# The symbiotic star CH Cygni

## II. The broad Lyα emission line explained by shocks

M. Contini[1,2], R. Angeloni[1,2] and P. Rafanelli[1]

[1] Dipartimento di Astronomia, University of Padova, Vicolo dell'Osservatorio 2, I-35122 Padova, Italy
e-mail: rodolfo.angeloni@unipd.it, piero.rafanelli@unipd.it
[2] School of Physics and Astronomy, Tel-Aviv University, Tel-Aviv, 69978 Israel
e-mail: contini@post.tau.ac.il

**ABSTRACT**

*Context.* In 1985, at the end of the active phase 1977-1986, a broad (4000 $km\,s^{-1}$) Lyα line appeared in the symbiotic system CH Cygni that had never been observed previously.
*Aims.* In this work we investigate the origin of this anomalous broad Lyα line.
*Methods.* We suggest a new interpretation of the broad Lyα based on the theory of charge transfer reactions between ambient hydrogen atoms and post-shock protons at a strong shock front.
*Results.* We have found that the broad Lyα line originated from the blast wave created by the outburst, while the contemporary optical and UV lines arose from the nebula downstream of the expanding shock in the colliding wind scenario.

## 1. Introduction

Symbiotic stars (SSs) were introduced as spectroscopically peculiar objects by Merril (1919). Nowadays they are interpreted as binary systems composed of a compact hot star, generally a white dwarf (WD), and a cool giant, whose mutual interaction via accretion processes is at the origin of complex spectra recorded at all wavelengths, from radio to X-rays.

The first models attempted to explain the symbiotic spectra in the light of a binary star configuration through the nebular emission arising from the photoionized portion of the cool giant wind. The photoionizing source was played by the WD alone. This simple scenario was fairly able to explain the first optical and radio data, but became inadequate for the detailed modeling of line emission (Nussbaumer & Vogel 1990) and for observations in new spectral ranges (Li & Leahy 1997). The observational evidence that actually both stars lose mass through important winds (Nussbaumer et al 1995) led to the colliding wind (CW) models (Girard & Willson 1987, Li et al. 1998, Kenny & Taylor 2005). A complex network of shock waves throughout the symbiotic systems could finally explain the observations, particularly in the high energy spectral ranges. The coupled effect of shocks and photoionization was therefore accounted for in the modeling of the line and continuum spectra (e.g. Contini 1997). Moreover, the network of gas and dust nebulae within specific SSs could be distinguished by composite models (Angeloni at al. 2007 a,b,c).

CH Cygni (CH Cyg) is amongst the most studied yet still discussed SSs. In a recent paper we have attempted a comprehensive and self-consistent modeling of continuum and line spectra in different epochs within the CW framework (Contini et al 2007, hereafter Paper I).

In particular, in 1977, CH Cyg underwent a powerful outburst that lasted until 1986. Towards the end, bipolar radio and optical jets appeared (Solf 1987) and spectral variations were observed in the UV and optical ranges. A remarkable characteristic of the 1985 spectrum was the appearance of a broad, strong Lyα emission line (Fig. 1), never evident in previous spectra (Selvelli & Hack 1985). This very peculiar spectral feature was, however, disregarded in Paper I, because it could not be explained by the CW model alone. The overall model of CH Cyg has been amplified and is presented in this paper.

We discuss the origin of the Lyα line appearance in 1985 and suggest an alternative interpretation accounting for shock waves that accompany the WD outburst during the active phase. In Sect. 2 we briefly review the observational trend and the previous theories suggested to explain the composite Lyα line profile. In Sect. 3 we describe the colliding-wind scenario and the resulting picture at the end of the 1977-1986 active phase (Sect. 3.1). Consequently, adopting Chevalier's (1982) theory on the interaction of the outburst with the circumstellar medium, we apply the results by Heng & Sunyaev (2007, hereafter HS07) about the Lyα broad line formation at high-velocity shock fronts (Sect. 3.2). A discussion appears in Sect. 4, while concluding remarks follow in Sect. 5.

## 2. The broad Lyα line: observations and former interpretations

During the outburst started in 1977, no Lyα emission from CH Cyg was recorded. Then, towards the end of the active phase (January 1985), a strong and broad Lyα emission line, as wide as 20 Å, appeared (Figs. 1 and 2), with a strong redward-shifted wing, suggesting expansion velocities up to 4000 $km\,s^{-1}$. This outstanding behaviour has not been repeated during the sub-sequent evolution throughout the active and quiescent phases. For instance, during the 1990-91 and 1995 quiescent stages, the Lyα virtually disappeared, together with the hot UV and optical continuum, while during the active phase 1992-1995 the Lyα line was similar in profile (Fig. 2), but 2-3 times weaker than in 1985 (Skopal et al 1998).

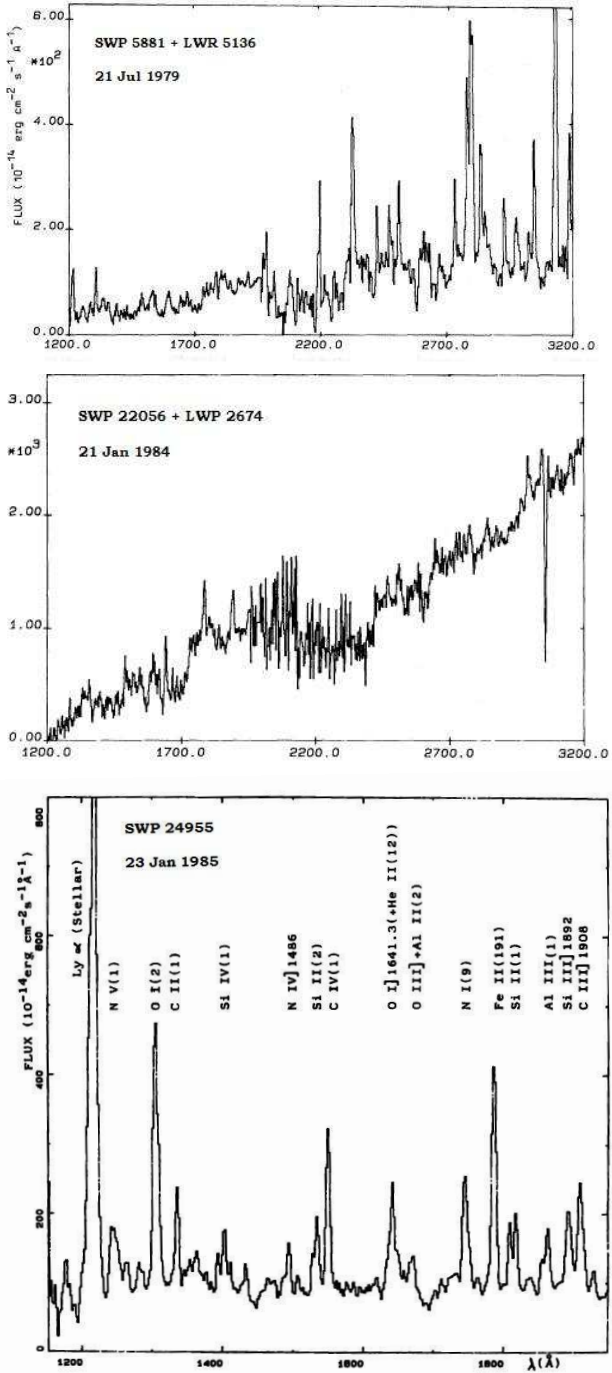

**Fig. 1.** IUE spectra of CH Cyg, showing the appearance of a strong and unusually broad Lyα in 1985, at the end of the 1977-86 active phase (adapted from: MSH88, top and central panels - Selvelli & Hack 1985, bottom panel)

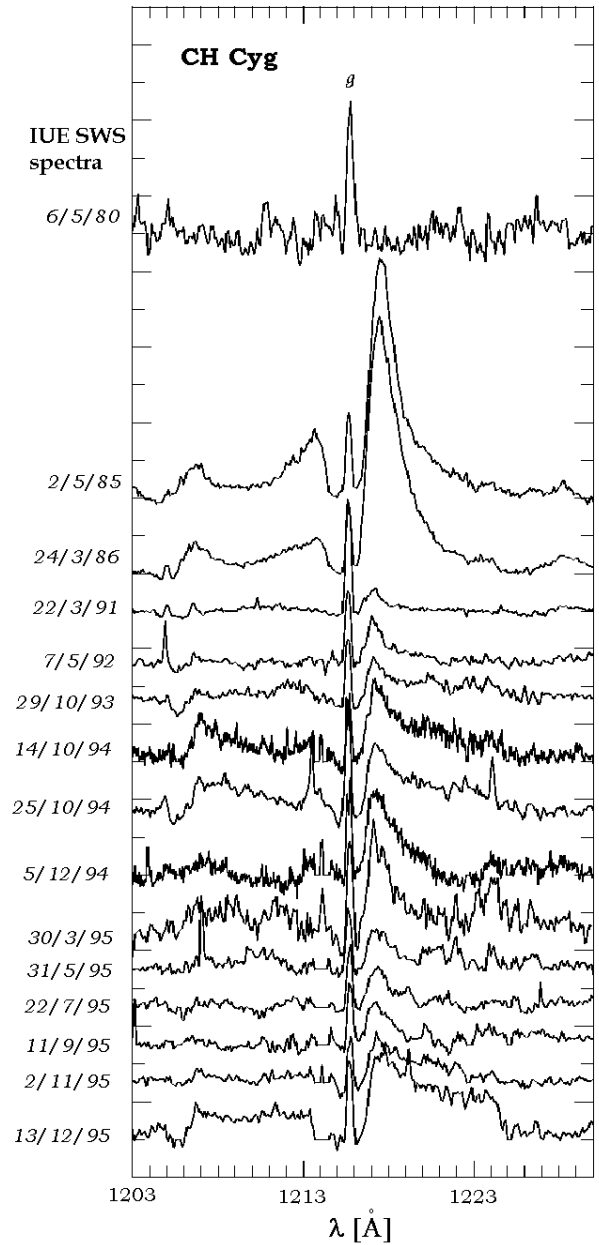

**Fig. 2.** Evolution of the Lyα profile in the high-resolution IUE spectra, from 1980 to 1995. On the left, the observation date are indicated; the $g$ marks the position of the narrow Lyα geocoronal emission (adapted from Skopal et al. 1998).

The standard Lyα profile has been decomposed into two - stable and variable - components (Skopal et al. 1998). While the former emission has usually been attributed either to the atmosphere of the red giant (Selvelli & Hack 1985) or to some circumbinary material (Skopal et al. 1998), the latter, i.e. the variable part of the profile that extends up to a high velocity (~ 2200 $km\,s^{-1}$), has been explained as a fast outflow in the vicinity of the hot star. An asymmetric high-velocity outflow (2000 $km\,s^{-1}$) was also indicated by the broad Balmer lines occurring in the spectrum on short time scales (Iijima et al. 1994). Such velocities and variabilities are similar to those characteristic of the shock between the stars created by collision of the winds (Paper I). Therefore, they are less anomalous than the broad (4000 $km\,s^{-1}$) Lyα line observed at the end of the 1977-1986 active phase.

A mechanism for explaining this broadening was proposed by MSH88 who claimed that, according to Johansson & Jordan

(1984), the Lyα line with an enhanced red wing, typical of line formation in an acceleration outflow, may be broadened by scattering by a high opacity. However, a stellar origin is difficult to reconcile with the sudden appearance of the broad Lyα line. MSH88 then proposed a formation region displaced from the orbital plane, connected with outflowing material, while the Balmer lines might originate in an accretion disk.

## 3. Theoretical scenarios

We would like to explain the exceptionally broad Lyα line emission in the frame of the shock-front network in CH Cyg. The results of the wind-collision model (Paper I) led to a detailed physical and morphological picture of the emitting nebulae within the SS. In particular it was demonstrated that the spectra depend on the system phase. In January 1985, at the end of the active phase that started in 1977, the UV and optical line ratios revealed that we were facing the circumstellar medium of the WD opposite the red giant. In this region, the dynamical consequences of the WD outburst can be compared with those of a supernova explosion, though on a different scale. Following Chevalier (1982), 8 years after the burst, the expanding blast wave reached a relative large radius and the velocity of the shock front is in the range of those characteristic of broad Lyα line emission by charge-transfer reactions (HS07).

### 3.1. Results of the colliding-wind model

Collision of the winds (Girard & Willson 1987, Kenny & Taylor 2005) from the two-component stars leads to two main shocks: the head-on shock between the stars, facing the WD, and the head-on-back shock expanding from the system outwards (mentioned in Paper I as the *reverse* and *expanding* shocks, respectively).

According to the colliding wind model, during the active phase, the broad Hβ lines with FW0M of 400-1200 $km\,s^{-1}$ were emitted downstream of the head-on shock between the stars (Angeloni et al. 2007a). Although this shock is actually a standing shock, it may be accelerated throughout the decreasing density of the atmosphere by the massive wind from the red giant. The broad Balmer lines emitted downstream are particularly strong because the intensity of permitted lines depends on the temperature of the star $T_*$ and on the ionization coefficient $U$. $T_*$ can reach more than 100,000 K during the outburst and $U$ is relatively high ($> 1$) because the emitting gas is close to the hot star. The broad lines decline both in intensity and width during quiescence because $T_*$ becomes $\leq 30,000$ K and the high velocity wind slows down. In a shock dominated regime, Rayleigh-Taylor (R-T), Richtmyer-Meshkov (R-M), and Kelvin-Helmholtz (K-H) instabilities at the shock fronts lead to fragmentation. Adopting a filling factor between 0.001 and 1 (Paper I), a maximum geometrical thickness of the filaments $\sim 10^{14}$ cm, and an average velocity of 600 $km\,s^{-1}$ during outburst, the variability time scale is between $\leq$ 1 hour and $\leq$ 1 month. The time scale increases during quiescence depending on the velocity decrease.

The sudden appearance of the broad (4000 $km\,s^{-1}$) Lyα line looks quite anomalous. Such a broad line could not even come from the nebula downstream of the expanding shock, which corresponds to $V_s \leq 150\ km\,s^{-1}$ (Paper I). Interestingly, the broad Lyα line appeared at about the same time as the line spectrum observed after November 1984, when the outburst was almost over. The spectrum revealed [OIII] 4363 and [OIII] 5007 lines, as well as high ionization UV emission lines, e.g. NV, CIV, SiIV, HeII, OIII] (Selvelli & Hack 1985, MSH88). Forbidden optical lines cannot survive the densities of $10^8$-$10^9$ $cm^{-3}$ that are characteristic of the shock between the stars. They must have been emitted from the nebula downstream of the shock expanding outward throughout the external region of the system, which in accordance with the colliding wind model (Girard & Willson 1987), is located on the side of the WD circumstellar region opposite the red giant. Consequently, the Lyα line, appearing at this very epoch, was emitted from high-velocity gas outflowing throughout the extended hemispherical region on the side of the WD less disturbed by the dynamical effects of the red giant wind.

### 3.2. The blast wave from the outburst

We adopt MSH88 suggestion that the broad Lyα emission region is connected with the outflow material.

Following Chevalier (1982), we consider the interaction of the outburst with circumstellar matter on the assumption that is built up by a steady wind. If the ambient density is described by $\rho \propto r^{-s}$ ($\rho = 1.4\ m_H\ n$, where n is the density in number $cm^{-3}$ and $m_H$ the mass of the H atom), the steady wind corresponds to s=2. The interaction of the freely expanding matter with the surrounding medium gives rise to a high-energy density region bounded by shock waves. Two shock fronts develop, one proceeding inward in the high-density region, the other expanding outward in the circumstellar medium.

The case of uniform expansion gas is described by s=2 and $\gamma$=5/3. For s=2 the radius of the outer shock is given by the Primakoff solutions: $R_{BW}=(3\ E/2\ \pi\ A)^{1/3}\ t^{2/3}$ (Chevalier 1982, Eq. 5), where E is the total energy, $\rho_0 = A\ R^{-2}$ ($\rho_0 = 1.4\ m_H\ n_0$, where $n_0$ is the pre-shock density of the gas upstream), and t is the time elapsed from the burst.

This equation is valid for times longer than the time of change $t_s = 0.677\ M^{3/2}/A\ E^{1/2}$, between that of the interaction of freely expanding matter with the surrounding medium and the following one, i.e. when the flow tends toward the self-similar solution for a point explosion in a power-low density profile (Sedov 1959). Here M stands for the ejected mass.

The velocity $V_s$ = dR/dt is constrained by the observed FW0M of the Lyα line 8 years after the burst. After some algebra we obtain $n_0$=1.45 $10^{-47}$ E. The mass ejected by the WD in the CH Cyg system during the 1977-86 outburst is about a few (2-3) $10^{-6}$ $M_\odot$ (Taylor & Seaquist 1985). The total energy is half thermal and half kinetic (Chevalier 1982). The high velocity observed after January 1985 presumes that the velocity of the ejecta was $\sim$ 14,000 $km\,s^{-1}$ at t=1 year. If all the ejecta had a velocity of 14,000 $km\,s^{-1}$, the associated kinetic energy would amount to 6 $10^{45}$ *erg*, the total energy E=1.2 $10^{46}$ *erg*, and consequently $n_0$ = 0.2 $cm^{-3}$. This in turn gives $R_{BW}$=2.25 $10^{17}$ *cm*.

To check this result we calculate for instance, $n_0$ close to the WD ($n_{0_{WD}}$) at a radius $R_{0_{WD}} = 2\ 10^{13}$ *cm*. We choose the distance that was found for the standing shock between the stars (Paper I) by modeling the continuum SED observed in May 1985. Although the interbinary region shows mixing of the winds from the stars, the standing shock facing the WD reflects the composition of the WD (e.g. Contini 1997). From $n_{0_{WD}} = n_{0_{BW}}\ R_{BW}^2/R_{0_{WD}}^2$ where $n_{0_{BW}} = 0.2\ cm^{-3}$, we find $n_{0_{WD}} = 2.5\ 10^7\ cm^{-3}$ in good agreement with the preshock density $n_0 = 5\ 10^7\ cm^{-3}$ adopted to explain the continuum SED at that epoch (Paper I, Table 1). Constraining $t_s$ within a maximum radius of $\sim$ 2. $10^{17}$ *cm* and adopting $n_0$=0.2 $cm^{-3}$, we find $t_s$= 5.2 years, confirming that Chevalier's model is valid in this case.

We can now adopt the theory of HS07, who treated the blast wave shocks in the SNR case. Their results show that broad Balmer and Lyman lines are produced by charge transfer reactions between the post-shock protons and the ambient atoms. A population of post-shock atoms follows with a broad velocity distribution (broad neutrals). For $V_s \geq 500$ $km\ s^{-1}$ the broad neutrals can produce Lyα that is blue- or red- shifted by resonance with the stationary atoms, hence providing an escaping way for the protons. For shocks with $V_s \geq 4000$ $km\ s^{-1}$, the luminosity ratio $\Gamma_{Ly\alpha/H\alpha}$ is $\geq 10$ (HS07 Fig. 1), so the Lyα line would be strong.

## 4. Discussion

As previously mentioned (Sect. 3.1), the broad Lyα appeared at about the same phase as the expanding shock within the colliding wind scenario. The radius of the expanding nebula $R_{exp} \sim 8\ 10^{16}$ $cm$ results from modeling the observed spectral energy distribution of the continuum (Paper I). Notice that the radius of the blast wave $R_{BW}$ calculated in Sect. 3.2, is about three times that of $R_{exp}$, while the physical parameters, $V_s=4000$ $km\ s^{-1}$ and $n_0=0.2$ $cm^{-3}$ of the blast wave, and $V_s=150$ $km\ s^{-1}$ and $n_0=10^5$ $cm^{-3}$ of the expanding shock, are very different. This is not surprising, considering that the blast wave stems from the outburst, while the expanding shock derives from the wind collision. The two shock fronts will hardly interfere even in the orbital plane, because R-T and K-H instabilities lead to fragmentation of matter at the shock fronts with filling factors < 0.01 (Paper I; Contini & Formiggini 2001).

A velocity of $\sim 4000$ $km\ s^{-1}$ was evident from the Lyα line profile after January 1985 in CH Cyg. The other UV and optical lines showed narrow FWHM profiles. In contrast, UV lines indicating expansions of 3000-4000 $km\ s^{-1}$ were observed after the 1985 eruption of the recurrent nova RS Ophiuci (Snijders 1987; Shore et al. 1996) and also optical lines after the last outburst in 2006 (Bode et al. 2007 and references therein) showed high-velocity expansion. The UV and optical lines strong enough to be observed can be emitted downstream of a strong shock if the preshock density is high enough to speed up the cooling rate ($\propto n^2$) downstream. Consequently the temperature will drop below $10^6$ K, leading to recombination. However, the densities are constrained by the critical densities for collisional deexcitation of the different ions, which are relatively low (< $10^6$ $cm^{-3}$) for forbidden lines. In order to emit strong enough forbidden lines (e.g. [OIII]), a shock with $V_s=4000$ $km\ s^{-1}$ should propagate throughout a medium with $n_0 \leq 10^4$ $cm^{-3}$ The lines are emitted from the gas beyond the temperature drop characteristic of shock-dominated regimes downstream, at a distance from the shock front of $\geq 7\ 10^{16}$ $cm$. Lower densities correspond to stronger forbidden lines at larger distances from the shock front. Strong [FeVII]6087, [FeX] 6375, and even [FeXI] 6986 are predicted at such high $V_s$. The UV lines are generally permitted or semi-forbidden, therefore they can also be emitted at higher densities. The OVI 1034 and CIV 1500 lines will be particularly strong. These predictions imply a preshock magnetic field of $B_0=10^{-3}$ Gauss.

Such densities are not unusual in the circumstellar medium, as they were found by modeling the [OII] and [NII] lines during the active phase 1998-2001 of CH Cyg (Paper I). The high preshock density would indicate that the shock is interacting with mass lost by either the progenitor star or by the SS in a previous burst. However, such high velocities were never seen in the line profiles of CH Cyg, except for the broad Lyα in 1985.

Interestingly, the energies calculated from the temperature ($\sim 2.94\ 10^9$ K) downstream of this strong shock front correspond to $\sim 250$ keV, not far from the gamma-ray energy range, while energies of $\sim 21$ keV ($2.4\ 10^8$ K) correspond to $V_s=4000$ $km\ s^{-1}$. We suggest that the observations of symbiotic star outbursts beyond hard X-rays might lead to interesting results.

The escape velocity of the WD is $v_{esc} = (2\ G\ M_{WD}/R_{WD})^{1/2} \sim 1.6\ 10^{13}/R_{WD}^{1/2}$, adopting a mass $M_{WD} \sim 1 M_\odot$. We recover a velocity of 14,000 $km\ s^{-1}$ similar to what is predicted at early times for the Lyα line FW0M profile, adopting a WD radius $R_{WD} \sim 1.36\ 10^8$ $cm$ in the CH Cygni system.

## 5. Concluding remarks

In 1985, at the end of the active phase 1977-1986, a broad (4000 $km\ s^{-1}$) Lyα line was observed that had never been present in previous spectra from the symbiotic system CH Cygni.

We have noticed that the broad Lyα line appeared contemporaneously with the optical-UV spectrum emitted downstream of the shock created by collision of the winds from the stars. This shock expands throughout the extended circumbinary region located on the side of the WD circumstellar region opposite to the red giant; as a result, the Lyα line originates somewhere in the hemispherical region opposite the WD, where the dynamical consequences of the burst are less affected by the red giant wind. The overall situation is similar to that of an SN explosion. This suggests that the broad Lyα emission line and the other optical and UV lines observed at the same phase could be emitted from different shocked nebulae. We consider that the broad Lyα line stems from the WD outburst.

Applying the theory developed by Chevalier (1982) for Type II supernovae to the interaction of the WD outburst with circumstellar matter, we have found that the expanding blast wave had reached a radius $R_{BW} \sim 2.25\ 10^{17}$ $cm$ 8 years after the burst for a shock velocity of 4000 $km\ s^{-1}$. We then applied the theory developed by HS07 for high-velocity shock fronts in SNR, namely, the broad Lyα line is produced by charge transfer reactions between the blast wave post-shock protons and the ambient preshock atoms. For shocks with $V_s \geq 4000$ $km\ s^{-1}$, the luminosity ratio $\Gamma_{Ly\alpha/H\alpha}$ is $\geq 10$ (HS07), so the observed Lyα line is strong.

The energy involved with the outburst is $E \sim 1.2\ 10^{46}$ $erg$, and the ambient density consistent with a velocity of 4000 $km\ s^{-1}$ is $n_0 \sim 0.2$ $cm^{-3}$. Higher velocities of about 14,000 $km\ s^{-1}$ predicted by the Sedov solution at early times, lead to temperatures in the downstream region of such a strong-shock front, corresponding to emission in the near gamma-ray frequency range. Such emission was not observed at that time, first because of technical inadequacy and also because, during the active phase, the WD circumstellar region opposite the red giant only beaome visible in 1985 when the broad Lyα appeared. The velocities had already decreased to about 4000 $km\ s^{-1}$. Actually, the SEDs in the VBU range during the 1977-86 period, presented in Paper I (Fig. 6), show that the black body flux from the hot star does not appear.

According to previous results (Paper I), the hard X-rays are emitted from a small region between the two component stars, while the soft X-ray are emitted from the extended circumbinary region. The results of this paper suggest that the hardest radiation ($\sim 250$ keV) comes from the WD circumstellar region close to the WD on the opposite side of the giant star, while hard X-rays could also be observed at distances $\leq 0.08$ pc from CH Cyg.

Finally, the complex light curves of CH Cyg (Eyres et al. 2002) show that the nebulae created by collision of the winds

Contini et al.: Broad Lyα emission line from CH Cygni 5and the dusty shells ejected from the red giant, expanding outward beyond the binary system, lead to temporary obscuration (Paper I). We suggest that the shock front at a relatively large radius corresponding to the blast wave may also contribute to obscuration episodes.

*Acknowledgements.* We are very grateful to the referee for constructive criticism and to Dina Prialnik for many helpful comments. We also thank P. Selvelli and A. Skopal for the kind permission to include some figures adapted from their papers.**References**

Angeloni, R., Contini, M., Ciroi, S., & Rafanelli, P. 2007a, AJ, 134, 205
Angeloni, R., Contini, M., Ciroi, S., & Rafanelli, P. 2007b, A&A, 471, 825
Angeloni, R., Contini, M., Ciroi, S., & Rafanelli, P. 2007c, A&A, 472, 497
Bode, M.F. et al. 2007, ApJ, 665, L63
Chevalier, R. A. 1982, ApJL, 259, L85
Contini, M. 1997, ApJ, 483, 887
Contini, M. & Formiggini, L. 2001 A&A, 375, 579
Contini, M., Angeloni, R., & Rafanelli, P. 2007, A&A, submitted, Paper I
Eyres, S. P. S., et al. 2002, MNRAS, 335, 526
Girard, T., Willson, L.A. 1987, A&A, 183, 247
Heng, K., & Sunyaev, R. 2007, ArXiv e-prints, 710, arXiv:0710.4282
Iijima, T., Strafella, F., Sabbadin, F., & Bianchini, A. 1994, A&A, 283, 919
Johansson, S. & Jordan, C. 1984, MNRAS, 210, 239
Kenny, H.T. & Taylor, A.R. 2005, ApJ, 619, 527
Li, P. S., & Leahy, D. A. 1997, ApJ, 484, 424
Li, P. S., Thronson, H. A., & Kwok, S. 1998, 1997 Pacific Rim Conference on Stellar Astrophysics, 138, 191
Merrill, P. W. 1919, PASP, 31, 305
Mikolajewska, J., Selvelli, P.L., & Hack, M. 1988, A&A 198, 150
Nussbaumer, H., & Vogel, M. 1990, Astronomische Gesellschaft Abstract Series, 4, 19
Nussbaumer, H., Schmutz, W., & Vogel, M. 1995, A&A, 293, L13
Sedov, L.I. 1959, Similarity and Dimensional Methods (New York:Academic Press)
Selvelli, P. L., & Hack, M. 1985, Recent Results on Cataclysmic Variables. The Importance of IUE and Exosat Results on Cataclysmic Variables and Low-Mass X-Ray Binaries, 236, 207
Shore, S. N., Kenyon, S. J., Starrfield, S., & Sonneborn, G. 1996, ApJ, 456, 717
Skopal, A., Bode, M. F., Lloyd, H. M., & Drechsel, H. 1998, A&A, 331, 224
Snijders, M. A. J. 1987, Ap&SS, 130, 243
Solf, J. 1987, A&A, 180, 207
Taylor, A.R. & Seaquist, E.R. 1985, IAU circular No. 4055